# Rapid and Inexpensive Reconstruction of 3D Structures for Micro-Objects Using Common Optical Microscopy
# (*Supplementary Part*)

Viatcheslav Berejnov

*Department of Mechanical Engineering and Institute for Integrated Energy Systems, University of Victoria, Victoria, V8W 3P6, Canada*

Fax: +1 250 721 6323;
Tel: +1 250 472 4202;
E-mail: berejnov@uvic.ca

**Abstract**
A simple method of constructing the 3D surface of non-transparent micro-objects by extending the depth-of-field on the whole attainable surface is presented. The series of images of a sample are recorded by the sequential movement of the sample with respect to the microscope focus. The portions of the surface of the sample appear in focus in the different images in the series. The indexed series of the in-focus portions of the sample surface is combined in one sharp 2D image and interpolated into the 3D surface representing the surface of an original micro-object. For an image acquisition and processing we use a conventional upright stage microscope that is operated manually, the inexpensive Helicon Focus software, and the open source MeshLab software. Three objects were tested: an inclined flat glass slide with an imprinted 10 µm calibration grid, a regular metal 100x100 per inch mesh, and a highly irregular surface of a material known as a porous electrode used in polyelectrolyte fuel cells. The accuracy of the reconstruction of the image features was found to be ~ 1% for the heights of ~ 100 µm with the corresponding image acquisition time being ~ 10 min.

**Introduction**

The microscopic structural analysis of a surface of non-transparent micro-objects never acquires the whole profile of an object in a single sharp focus. When the profile of the object covers more than the attainable depth of focus provided by the particular objective, the part of the object surface that is outside of the focus looks blurry and thus cannot be measured and analyzed. The thickness, $d$, of an imaginary layer containing the portion of the object surface that looks sharp is called the depth-of-field and is a function of the lenses in the objective. The effect of surface defocusing of the non-transparent objects becomes even worse when we increase the magnification of the microscope objective, expecting to capture the small details of the surface more precisely. Since the magnification of the lenses of the objective, $Ma$, is proportional to the objective numerical aperture, $Na$, and the $Na$ increases as the focal plane approaches closer to the objective, then for the first order of estimation [1, 2] the depth of field decreases as:

$$d \approx \frac{\lambda}{Na^2},$$

where is a wavelength of light that can be taken for an estimate as ~ 0.65 μm.

This effect of defocusing the micro-object surface seriously limits those benefits of a visual control that are primarily originated by the invasive nature of an optic examination. The research and development of the modern hydrogen fuel cells have met a similar defocusing problem at the same point, experiencing a significant need in the simple and invasive method of the surface control for the hi-tech material called the carbon paper [3]. The surface of the carbon paper, also know as a porous electrode, is non-transparent, rough, and well microstructured [4]. The carbon paper is generally manufactured from the carbon filaments assembling them in the irregular 3D network and can vary as much as 100 μm in height, having the features of 10 μm size in the X-Y plane.

There is an approach that allows overcoming the defocusing problem and observing the whole surface of the micro-object in a single focus by constructing a synthetic image of that object. Generally, the method is taking the multiple images (slices) corresponding to the different positions of the focus and selecting from each optical slice a portion of the object surface that is in focus, Figure 1. The combination of those in-focus portions into a single image will provide a new, synthetic 2D image of the original surface that is sharp everywhere. In other words, this method allows extending the depth of field, $d$, over the whole attainable surface of the object. Next, since every in-focus portion is indexed by the appropriate slice, and the distance between slices is known, it is possible to interpolate those assembled portions into a single 3D surface that will represent an approximation of the surface of the original object.

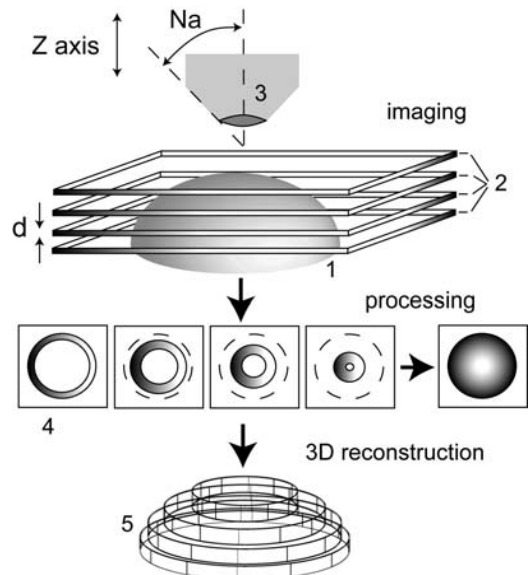

**Figure 1** Schematic representation of an idea of the surface reconstruction due to the fusion of the multiple in-focus images. The sample of the micro-object (1) is optically sliced (2), advancing the objective (3) of the numerical aperture, $Na$, along the vertical Z axis. Every slice of the focal plane (2) has its depth of focus, $d$, providing an appearance of the focused parts on the stack of the recorded images (4), with respect to the Z-indexes of the images, while the rest of every image is defocused. Combining the focused parts (shadowed) in a recorded series (4), it is possible to construct the 3D surface (5) of the original sample (1).

The first real challenge here is to select the sharp portion of an image that is in focus, indexing this portion to the given focal plane. The second challenge is developing the method of combining or fusing these portions into one single 2D image providing the new, synthetic depth of field. Numerous works have been done in attempt to solve these challenges by applying different methods of image filtering and fusing of the filtered focused portion in a resultant image [5-10]. Some of these methods are implemented in the commercially available software products [11-13], which currently are in very different categories of the budget rate $$$-$$'$$$ and could be used with or without supplemental optical hardware.

In this article we present the results of testing one example of these commercially available products: the Helicon Focus software [13] from the lowest budget rate $$$. Using Helicon Focus we reconstruct the 3D surface of the three very different microscopic samples: the optical calibration standard, the regular square mesh, and the microstructured irregular network of the carbon filaments. For the multi-focus image acquisition we use a convenient upright stage microscope in the bright light mode with the

**Experimental and Method**

In the experiments we used three different objects that had microstructured surfaces: an optical calibration standard from the Leica Company, Germany, with the 2 mm grid scale and 10 µm tick's resolution, the regular square mesh of the metal wires, and the porous layered material made from the carbon fibers. The metal mesh was fabricated from the copper wires with 100x100 meshes per inch and the wire diameter of 0.0114 cm (0.0045''). The porous layer was the Toray carbon paper B-2/060/40WP from E-TEK Company, USA, the thickness of the sheet of this material was ~ 200 µm and the diameter of the carbon filament was about 8 µm.

The samples 2.5x2.5 cm of the copper mesh and carbon paper were mounted horizontally on the table of the Leica DM LM microscope used in the inverted mode. The resolution of positioning in the X-Y plain is 0.1 mm. The focal wheel of the microscope's vertical stage (Z axis) has two modes of positioning with 1 µm and 4 µm intervals, respectively. The mercury lamp was used for the light source. We used the four Leica objectives: 5x/0.15, 10x/0.30, 20x/0.40, and 50x/0.50, where the first number is the magnification, $Ma$, and the manually operated Z axis stage. We found that if the moving interval of the focal plane in the Z direction is ~ 8 µm, then Helicon Focus could recover the height of the microscopic features with ~ 1-2% accuracy and the manual image acquisition takes ~ 10-12 min. The accuracy could be further improved if an interval of the multi-focus imaging decreases to 4 µm that leads to ~ 25 min of image acquisition time. Further improvement of accuracy corresponding to 1 µm of the microscope stage displacement in the Z direction leads to the non-practical acquisition time of ~ 1.5 h.

| Objective | $Ma$ | $Na$ | $d$, µm |
|---|---|---|---|
| HC PL Fluotar | 5x | 0.15 | 29 |
| HC PL Fluotar | 10x | 0.30 | 7.2 |
| N Plan | 20x | 0.40 | 4.1 |
| N Plan | 50x | 0.50 | 2.6 |

**Table 1** Parameters of the objectives.

second one is the numerical aperture, $Na$ (see Table 1 for details). The 8-bit images were captured with the black/white camera Retiga 1300i and preprocessed with the software QCapture Pro 6.0 from QCapture Inc.

In our measurements the sample was mounted on the horizontal microscope stage. The stage was manually advanced to the objective direction with one of the following intervals: 1 µm, 4µm, or 8 µm. Every interval was applied in a sequence eventually attaining the whole surface of the sample. The QCapture Pro software recorded a JPG image at every advancing step and eventually collected a whole series of recorded images. The stack of images indexed by the Z coordinate was processed with the Helicon Focus software.

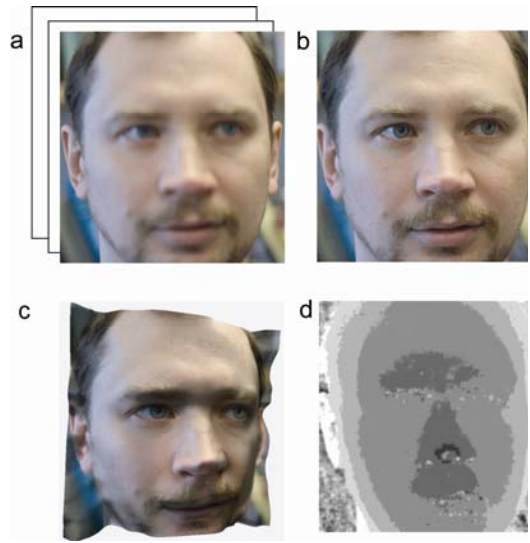

**Figure 2** Processing the stack of the images by the Helicon Focus software. (a) is a stack of the acquired images with different focusing; (b) is the synthetic image that is sharp everywhere; (c) is the 3D approximated surface of the depth map (d)

*Image processing by Helicon Focus*: Accordingly to the Helicon Focus (HF) company the HF software can provide two different algorithms of reconstruction abbreviated as A and B.

Following the A method, the software analyzes the gradient of the pixel intensity inside of the small pre-set area of the radius, $R$, (where $R$ is a setting parameter) assuming that the higher gradient corresponds to the better focused portion on the image. Then, every pixel on the image is multiplied by the weight coefficient calculated with respect to its nearby gradient. As a result, only those pixels that have the highest weights will impact the final image, referring to the fact that they are in the sharp portion of the particular optical slice of the stack; the depth map is calculated afterwards.

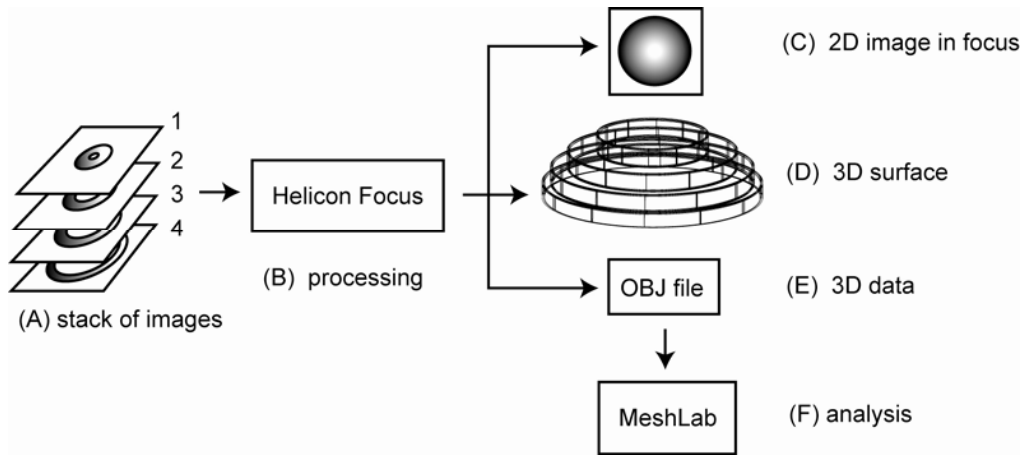

**Figure 3** The method of obtaining the 3D data of the object surface using the Helicon Focus and MeshLab software. For notation see the text.

Opposite to the A method, the second, B method uses the depth map as a direct tool for calculating the final image. The B method searches for the pixel with the nearby highest gradient. Than the HF program saves a portion of the image with respect to this gradient into the depth map flattening the color to the gray value from 0 to 255 accordingly to the index of the processed image in the original stack. This depth map is then smoothed and filtered in order to remove the noise pixels and the clusters of the pixels unrelated to the reconstruction. Then the HF program creates the final image substituting the pixel from the depth map by the pixel from the appropriate image in the stack.

Eventually, both methods end up with the final synthetic image, Figure 3 (C), containing the maximal amount of the sharp portions of the source images.

The depth map of both methods is used by the HF program for constructing the 3D surface of the original object, Figure 3 (D). The 3D surface processed in this manner can be rendered in the 3D space and saved as an OBJ file (E) [14], which is standard of the Wavefront Technology for definition and transferring the geometrical properties of the 3D objects. This OBJ file is compatible with the open source MeshLab software (F) [15]. We used the MeshLab program for generating the profiles of the 3D surface sectioned by the Z-X or Z-Y planes.

**Results**

The linear scale in the X-Y plane was calibrated with the 2 mm optical calibration standard, having 10 µm resolution.

Recording this calibration standard with the different objectives provided the series of X-Y scaling factors we used for the X-Y in-plane image processing. Calibration in the Z axis is less obvious because it includes multiple stages of the image processing, requiring different software and algorithms. The general idea here is to measure and compare heights and patterns of test objects in two ways: directly and with help of the 3D reconstruction techniques.

*Optical 10 µm calibration standard*: The glass slide containing the calibration standard grid was inclined at $7.01+/-0.01°$ on the horizontal microscope stage and its images were recorded for the different focal distances attaining the all scale that was fitted in the CCD camera field of view, Figure 4. We used the 10x/0.30 objective for these measurements. The stack of the inclined standard images was processed using the HF software and the 3D surface was obtained, Figure 4.

Because of the relatively small quantity of the multi-focus images, in Figure 4 (d) we see the stair-like distortion of the otherwise flat standard that does not affect the Z axis calibration but does affect the perception of the sample shape. The highest elevation on the image presented in Figure 4 (d) is 13 steps (meaning that the 13 images of different focuses were included in the analysis), where the each step is the upright lift of the microscope stage in the Z direction. Every lift corresponds to the two ticks on the scale of the focus wheel where every tick is 4 µm. Thus, the total elevation of the inclined optical calibration standard slide measured from Figure 4 (d) is 104 µm. Since we know the inclination and the length of the optical calibration standard that is fitted on the image we could calculate this elevation directly and it gives us 102.5 µm. The difference between the direct measuring and using the Helicon Focus software is ~ 1.4%.

*Metal mesh*: The sample of the copper mesh was mounted horizontally on the microscope stage and a series of the mesh images was collected with the 10x/0.30 objective. This image stack was processed with the HF software in a similar way as it was done in the previous example. The 2D image of the extended field of depth presented in Figure 5

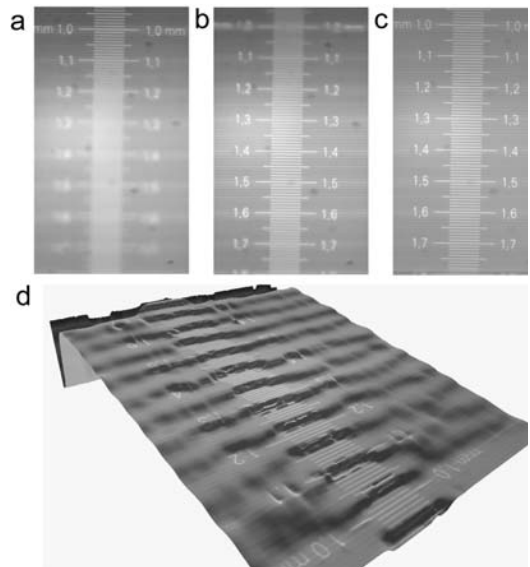

**Figure 4** The inclined glass slide with the calibration standard grid recorded with the 10x/0.30 objective. The panels (a) and (b) show the optical calibration standard at Z=0 µm and Z=56 µm, respectively; the inclination is 7.01 degrees. The processed image is presented on (c) and the panel (d) shows the 3D reconstructed surface of the image (c) using the HF software.

(a) and the 3D surface of the mesh sample presented in Figure 5 (b) were reconstructed.

Then the single wire of mesh was released non-deformed from the mesh network. This single wire was positioned on the microscope stage in order to provide in the field of view the maximal depth for the wire pitches. The image stack of this mesh element was collected for multiple focuses and processed with the HF program, Figure 5 (c) and (d). The whole in-focus image, Figure 5 (c), was obtained and the 3D surface of the mesh element was formatted as the OBJ file.

The OBJ data was processed with the MeshLab software and the surface of the wire was rendered in the 3D space. Then the 3D surface was vertically sectioned along the length of the wire, see the black curve in Figure 5 (d), and the numerical profile of this section was retrieved and saved by the MeshLab tools. The number of the steps on the profile representing the number of the images included for the HF analysis in the stack is 14. Every step corresponds to the vertical lift of the microscope stage on 8 µm. Thus, the gap for the black curve in Figure 5

(d) corresponding to the pitch of the mesh sample is 112 µm. The direct measurement of the same gap using the image of the mesh wire turned on 90 degrees gives 114 µm, Figure 4 (e). The difference between the direct measurement and the measurement including the consecutive application of the Helicon Focus and MeshLab software is ~ 2.6%

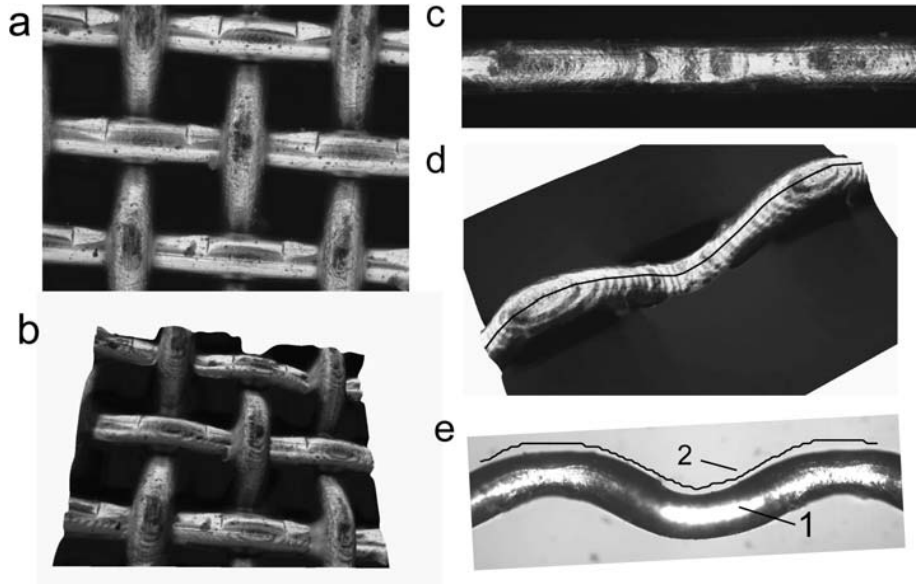

**Figure 5** Reconstruction of the 3D surface of the copper mesh (a) and (b) and calibration of the scale of the Z-axis using reconstruction of the single element of this mesh (c), (d), and (e). The panels (a) and (c) represent the top views of the mesh and wire after the image processing, respectively. The panels (b) and (c) show the 3D reconstructed profile, where the black curve on (c) depicts the cross section taken from the 3D-surface data and plotted on (e). The panel (e) presents a side view (1) of the segment shown on (c) and the profile data (2) taken from the image (d).

*Carbon paper*: The sample of carbon paper was mounted on the microscope stage horizontally. Then the series of the multi-focused images of the given view area were taken by varying the position of the stage along the Z axis within the 4 µm interval, *h*. Four different objectives were tested: 5x/0.15, 10x/0.30, 20x/0.40, and 50x/0.50. The stacks of the images indexed by the Z coordinate were processed with the HF software and 3D surfaces were created for every stack, Figures 6 and 7.

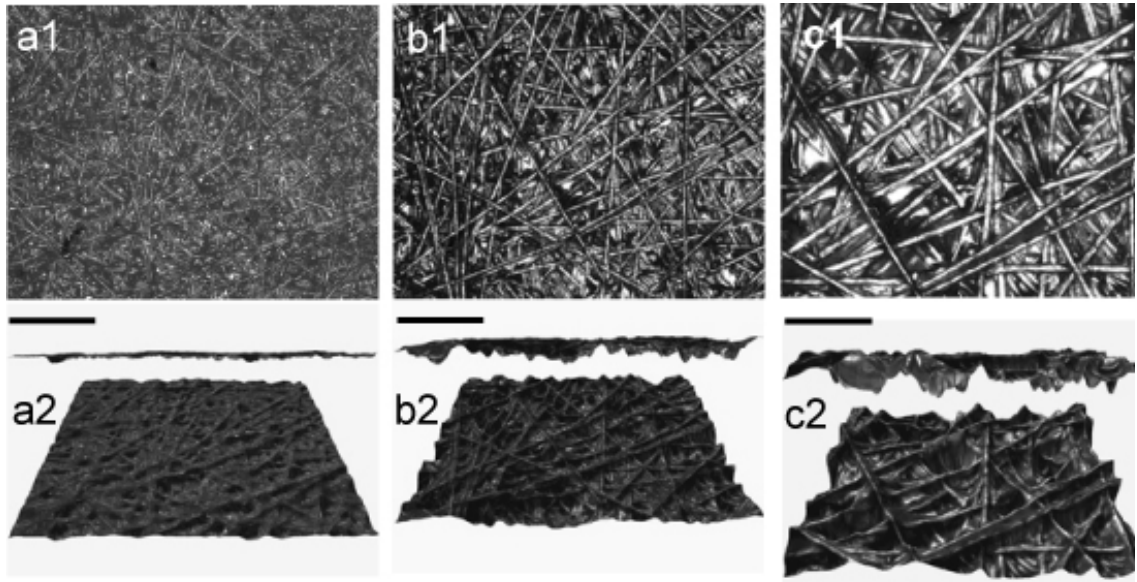

**Figure 6** The processed and reconstructed images indexed 1 and 2 respectively, for the three different objectives and the same view area. (a) is 5x/0.15 objective, the bar is 400 µm; (b) is 10x/0.30 objective, the bar is 200 µm; and (c) is 20x/0.40 objective, the bar is 100 µm. The 3D images are presented by the side and the perspective projections of the carbon paper sample

In Figure 6 we present the three tests with three different objectives acquiring and processing the same sample area. The values of depth of field, $d$, for those objectives are presented in Table 1. We can see a simple fact that if $d >> h$ (objective 5x/0.15), then the current method is able represent the very average 3D structure of the sample. In case if the interval of the stage lifting, $h$, is of order of $d$ (objective 10x/30), then the current method can reproduce the features of order of the size of the carbon fiber (~ 8 µm). For this condition the acquisition area is of order of 1 mm$^2$ that is sufficient for statistical analysis of the sample structure having the area of the features of order of 10-100 µm$^2$. The best resolution the method provides when $d$ is equal or less than $h$ (objectives 20x/0.40 and 50x/0.50, see Figure 7). There are two wells w1 and w2, the black point depicts the deepest depth of 120 µm that is located in the well w1.

Applying the high magnification objectives with the reasonable for the manual operation focus interval ~ 4 µm provides the best quality of the 3D reconstruction of the surface structure; with fields of view of 0.5 mm$^2$ and 0.2 mm$^2$, respectively. This aspect makes the highest magnification objective 50x/0.50 to be very convenient for inspecting the local features of the carbon paper sample, while the objectives 20x/0.40 is still good for acquiring the statistics of the surface structure.

**Conclusion**

Using the conventional manually operated microscope and the inexpensive software we developed a simple method for constructing the 3D surface of non-transparent micro-objects. The method consists of three major steps: *i)* acquiring the series of images of a sample with the sequential movement of the microscope stage, *ii)* applying the Helicon Focus software for the image stack and generating the synthetic sharp image and the 3D surface of the sample, and *iii)* analyzing the 3D sample surface with the MeshLab software. Three objects were tested: an inclined flat glass slide with an imprinted 10 µm calibration grid, a regular metal 100x100 per inch mesh, and a highly irregular surface of a material known as a porous electrode used in polyelectrolyte fuel cells.

We found this method to be very convenient for inspecting the microstructured surface of the carbon paper at variety of linear scales. We have demonstrated that for the carbon paper we were able to acquire both the large scale (~ 1 mm) structures suitable for statistical analysis of the "whole" sample as well as the small scale (~50 μm) features useful for characterization the "local" properties of the inspected material.

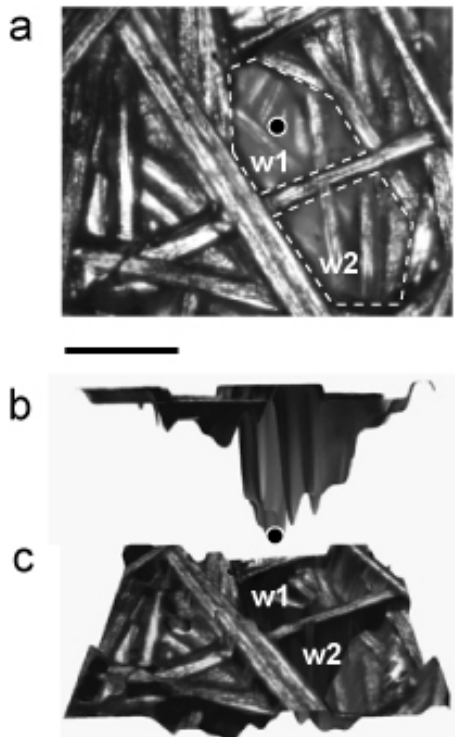

**Figure 7** The processed (a) and the 3D reconstructed images (b) and (c) for the side and for the perspective views, respectively; the objective is 50x/0.50. The bar is 50 μm.